\documentclass[twocolumn,prb,showpacs,floatfix]{revtex4}  

\usepackage{graphicx}
\usepackage{amssymb}

\begin{document}

\title{Energy landscape of relaxed amorphous silicon}

\author{Francis Valiquette and Normand Mousseau}

\affiliation{D\'epartement de physique and Centre de recherche en physique
et technologie des couches minces, Universit\'e de Montr\'eal,
C.P. 6128, succ. Centre-ville, Montr\'eal (Qu\'ebec) H3C 3J7, Canada}

\date{\today}

\begin{abstract}
We analyze the structure of the energy landscape of a well-relaxed
1000-atom model of amorphous silicon using the activation-relaxation
technique (ART nouveau).  Generating more than 40,000 events starting
from a single minimum, we find that activated mechanisms are local in
nature, that they are distributed uniformly throughout the model and
that the activation energy is limited by the cost of breaking one
bond, independently of the complexity of the mechanism. The overall
shape of the activation-energy-barrier distribution is also insensitive to
the exact details of the configuration, indicating that well-relaxed
configurations see essentially the same environment. These results
underscore the localized nature of relaxation in this material. 
\end{abstract}

\pacs{
61.43.Dq, 
66.30.Hs, 
02.70.-C 
 }

\maketitle

\section{Introduction}
\label{sec:intro}

The dynamics of many complex materials is dominated by activated
jumps over energy barriers generally higher than $k_BT$. For these
systems, the energy-landscape picture, which focuses on the
topological relation between locally stable states, has proven very
valuable.  Disconnectivity graphs, for example, introduced by
Czerminski and Elber~\cite{czerminski90} and applied extensively by
others,~\cite{becker1997, wales98,doye99a,doye99b,mortenson01} have
provided a first classification of the dynamics of complex systems
based on the structure of their respective energy landscape. 

In parallel with these developments, oriented towards reconstructing
the topology of the energy landscape, there has been some
efforts in trying to characterize the energetics and the nature of the
events in clusters,~\cite{wales98,doye99a,doye99b,malek00} proteins
~\cite{mousseau01,mortenson01,wei02a} and in amorphous materials.
~\cite{barkema98,mousseau00b,song00,middleton01} This extensive
sampling is essential in order to try to connect the properties of
the landscape with the dynamics measured experimentally, it also
serves to build a better understanding of the generic properties of
various networks: low vs. high connectivity, bulk vs. finite-size
systems, covalent vs. ionic bonding, etc. 

In this paper, we present an extensive study of the structure of the
energy landscape of amorphous silicon around two minima. Less
extensive studies of this material were already presented both by our
group~\cite{barkema98,mousseau00b,song00} and Middleton {\it et
  al.},~\cite{middleton01} using two different approaches. Using ART
nouveau, we generate more than 42 000 activated events around a
well-relaxed minimum and analyze the properties of the reaction paths
connecting this initial minimum to a nearby saddle point and new
minimum. We find that : (1) the activation mechanisms in {\it a}-Si
are local, limiting somewhat the usefulness of the configurational
energy landscape picture, (2) the activation energy is essentially
limited by the energy required to break a single bond, (3) the
entropic barrier is almost identical for all events, (4) more than 20
\% of all events generated are bond-switches of the
Wooten-Winer-Weaire type, and (5) the number of events seems to be of
the order 30 to 60 per atom.

\section{Details of the simulation}

We study here the energy landscape around two well-relaxed 1000-atom
configurations of {\it a}-Si. We focus mostly on the first one,
repeating the simulation on a second configuration only to ensure that
the results are generic and not dependent on some specific feature of
the configuration.

The energy model used is a modified version (mSW) of the empirical
Stillinger-Weber (SW) potential~\cite{stillinger85,vink01} developed
to ensure that the elastic and structural properties of the amorphous
phase of silicon are in agreement with experiment.~\cite{vink01} With
its original parameters, the SW potential describes both the liquid
and the crystalline phases with good accuracy but fails to reproduce
the experimental structure of the amorphous
phase.~\cite{ding86,cook93,barkema96} Recently, Vink and collaborators
proposed a slightly different set of parameters (see Table
~\ref{table:params}), which generates the right amorphous structure in
addition to providing the correct vibrational properties.  Because of
its empirical nature and the fact that it is not optimized for
dynamics, the energy barriers computed with this potential must be
taken with some care.  However, since the structural properties of
this material are well described by mSW, the qualitative features of
the energy landscape are expected to be correct.

\begin{table}
\caption{Parameters of the original Stillinger-Weber potential as well
  as those modified by Vink {\it et al.}, as described in
  Ref.~\protect\onlinecite{vink01}.  We use the latter set in this
  paper.}
\label{table:params}
\begin{tabular}{lll}
Parameter       & standard SW           & modified SW           \\
  \hline
  $\epsilon$ (eV) & 2.16826      & 1.64833       \\
  $A$             & 7.049556277  & 7.049556277   \\
  $B$             & 0.6022245584 & 0.6022245584  \\
  $\sigma$ (\AA)  & 2.0951       & 2.0951        \\
  $p$             & 4            & 4             \\
  $a$             & 1.80         & 1.80          \\
  $\lambda$       & 21.0         & 31.5          \\
  $\gamma$        & 1.20         & 1.20
\end{tabular}
\end{table}

\subsection{ART nouveau}

While our previous study of relaxation in {\it a}--Si used a version
of ART that could not identify saddle
points,~\cite{barkema96,mousseau98,song00} the results presented here
are obtained using ART nouveau, the latest version of the
activation-relaxation technique presented in
Ref.~\onlinecite{malek00,wei02a}.  Using ideas similar to those
proposed by Munro and Wales,~\cite{munro99} ART nouveau applies a
Lanczos scheme to compute directly the lowest second derivative of
the energy during the activation phase and ensures convergence to
the saddle point to any desired precision.

Events are generated in the following way: In order to leave the
harmonic well, one atom is selected at random. This atom and its
neighbors, contained in a shell of radius 3.5 \AA, are moved
iteratively in a randomly chosen direction, while keeping the energy,
projected in the perpendicular directions to a minimum.  At each step,
a Lanczos scheme is used to compute the lowest eigenvalues and
eigenvectors associated with the curvature of the energy
landscape. We consider that the configuration has left the harmonic
well when the lowest eigenvalue falls below $-50$ eV/\AA$^2$.

The activation process {\it per se} starts from this point. The
configuration is slowly pushed up along the direction of lowest
curvature until the modulus of the force falls below 0.5 eV/\AA ---
indicating that the configuration has reached a saddle point--- or
until the lowest eigenvalue become positive --- indicating that the
trajectory selected is back in the initial harmonic well region and
must be rejected.  After reaching the saddle point, the configuration
is pushed slightly away from it, and is relaxed into a new local
energy minimum, called the ```final minimum''. This event is stored
and a new event is started from the same initial minimum.

\subsection{Properties of the initial configurations}

\begin{figure}
\centerline{\includegraphics[width=8.50cm]{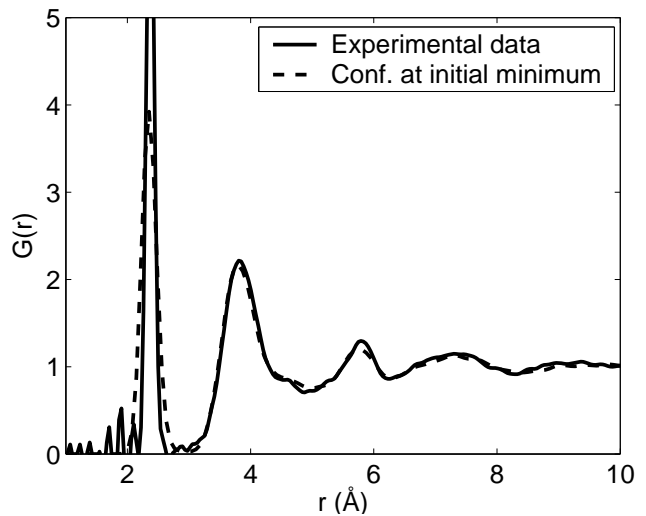}}
\caption{Solid line: radial distribution function (RDF) of the
  1000-atom {\it a}-Si model used in the simulation; dashes:
  experimental RDF measured by Laaziri {\it et
    al}\protect\cite{laaziri99}.}
\label{fig:rdf}
\end{figure}

The two initial configurations used here were prepared using ART
nouveau and mSW. Starting from a 1000-atom randomly packed unit cell
with periodic boundary conditions, ART events were applied until the
configurational energy equilibrated.  We used a Metropolis
accept/reject algorithm, based the energy difference between
consecutive local minima, as described in
Ref.~\onlinecite{barkema96,mousseau00a}.  The first configuration has an
energy per atom of $-4.000$ eV, with 20 five-fold and 26 three-fold
defects. The radial distribution function is in good agreement with
recent experimental data(Fig.~\ref{fig:rdf}). The model is therefore
of comparable quality to the models discussed in
Refs.~\onlinecite{mousseau98,barkema00}.  This configuration is used
as the origin of all events for the first 42~000-event run.

The second configuration was prepared by further applying ART on the
initial configuration, at a Metropolis temperature $T=0.25$ eV, for a
few thousands of events, until the average displacement per atom
reached 1 \AA. We generated more than 7000 events around this second
minimum in order to ensure that the features of the energy landscape
were statistically independent of the minimum selected. All the
quantities analyzed here are the same for both minimum, confirming
that the properties of the landscape are not affected by the details
of the topology. In view of the similarities between the properties of
these two sets of data, most of the discussion will focus on Minimum 1.

\subsection{Search for activated events}

All events presented here are characterized by three
configurations: the initial minimum (either Minimum 1 or 2), a
first-order saddle point and a second (final) minimum, obtained by
relaxing the configurational energy from the saddle point, in a
direction opposite to that of the initial minimum. As discussed in
Ref.~\onlinecite{malek00}, these events are reversible both from the
saddle point and the final minimum.

Each event is generated starting from the same initial minimum, but
with a different random initial direction, the rest of the procedure
being deterministic.  It takes about 400 force evaluations to
generated a single event. About one third of the random launches lead
to a trajectory that brings the configuration back in the harmonic
basin; these trajectories are simply rejected and a new random
direction selected. The whole search took about 4 weeks on a single
processor of a Regatta Power4 IBM.

\section{Results}
\label{sec:results}

\subsection{Distributions on events}

\begin{figure}
\centerline{\includegraphics[width=9.5cm]{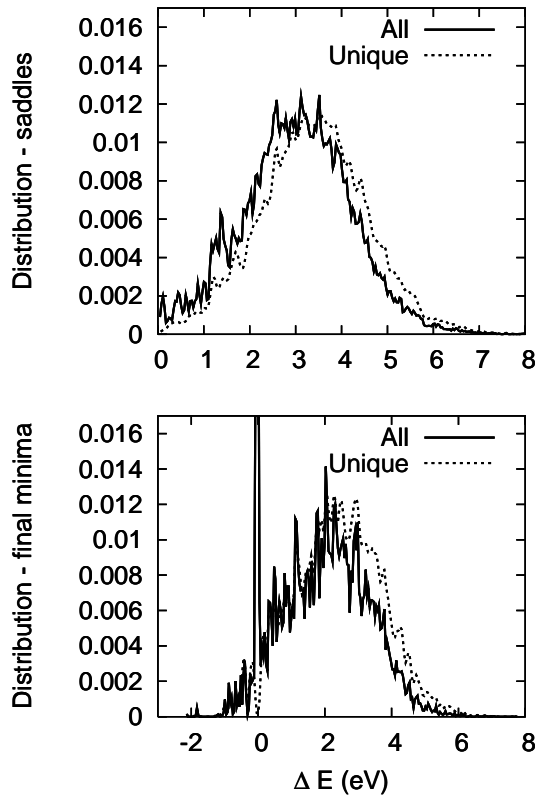}}
\caption{Solid lines : Normalized distribution of energies for the 42
  581 events generated around Minimum 1. Top: distribution of the
  activated energy $E_{\mathrm{saddle}} - E_{\mathrm{initial}}$. Bottom:
  distribution of the asymmetry energy $E_{\mathrm{final}} -
  E_{\mathrm{initial}}$.  Dotted lines : distribution of energies for
  the 8799 different (unique) saddle points (top) and the 6519
  different final configurations. }
\label{fig:distr}
\end{figure}

We first discuss the distributions of events generated from Minimum 1.
42 581 events were generated following the procedure discussed in the
previous section. The distribution of activated ($E_{\mathrm{saddle}}
- E_{\mathrm{initial}}$) and asymmetry ($E_{\mathrm{final}} -
E_{\mathrm{initial}}$) energies for all these events are given in
Fig. ~\ref{fig:distr}. The activated energies form a continuous
spectrum, from 0 to more than 7 eV, with an average barrier height of
3.0 eV and a distribution width of 1.2 eV.  Both the width and average
barrier energies are much lower than those reported earlier using the
previous version of ART.~\cite{mousseau00b} This underlines the limit
of the original ART, which cannot converge directly to a saddle
point. As shown in Ref.~\onlinecite{song00}, the error on the barrier
using the initial version of ART could be as high as 1 eV. As with the
method of Munro and Wales,~\cite{munro99} on the contrary, ART nouveau
makes it possible to identify the transition point to any desired
accuracy.  Comparing with experiment, we find that the average barrier
height is in overall agreement with measurements of Shin and
Atwater~\cite{shin93} which indicate activation barriers extending
from 0.25 eV to about 2.8. The typical activation energies we find
correspond also to the isothermal calorimetric data of Roorda {\it et
  al.} which indicate high-activation barriers.

Although there are no experimental information to compare with, it is
useful to examine the distribution of the asymmetry energy, i.e., the
energy difference between the final and initial minima (see the bottom
panel of Fig.~\ref{fig:distr}).  The average of the distribution is at
1.7 eV, with a width of 1.4 eV.  There are only a few events with a
final energy lower than the initial as should be expected, since the
initial configuration is already very well-relaxed. As shown below,
the distribution around Minimum 1 is essentially identical to that
around Minimum 2.  This results suggests that the overall shape of the
configurational energy landscape is not sensitive to the details of
the configuration, contrary to what one could think.

\begin{figure}
\centerline{\includegraphics[width=6.5cm]{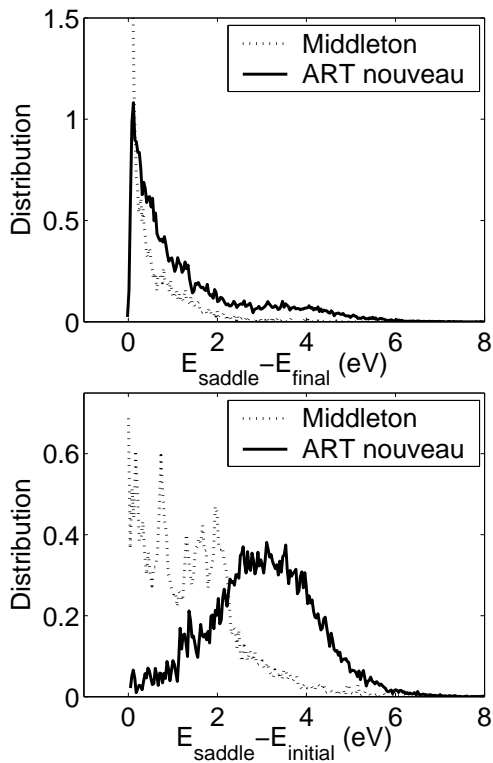}}

\caption{Comparison between the distributions of activation energy
  calculated from the final (top) and the initial (bottom) minimum
  for the set of events computed by Middleton and Wales
  ~\protect\cite{middleton01} and that generated with ART nouveau
  from minimum 1.}
\label{fig:wales}
\end{figure}

It is also useful to compare our results with those of Middleton and
Wales obtained on the same system using the eigenvector following BFGS
approach (EF-BFGS);~\cite{middleton01} ART nouveau is similar in
spirit to this method.  In Fig.~\ref{fig:wales}, we plot the
activated energy calculated from the {\it initial} minimum and the
{\it final} ($E_{\mathrm{saddle}} - E_{\mathrm{final}}$) for both sets
of events.  The barrier distribution calculated from the initial
minimum show that the sampling of events differs seriously for both
methods; while EF-BFGS seems to favor strongly events with a barrier
below 2 eV, ART selects events in a more Gaussian way. Since both
methods follow closely the direction corresponding to the lowest
eigenvalue to the saddle point, this difference is solely due to the
algorithm used to leave the harmonic well.  Remarkably,
however, this selection has very little impact on the barrier
distribution calculated from the {\it final} minimum; in this case,
both methods find the same barrier distribution, which is heavily
skewed towards very lower barriers. We discuss the significance of
these dissimilarities in Section~\ref{sec:discussion}.

\begin{figure}
\centerline{\includegraphics[width=6.5cm]{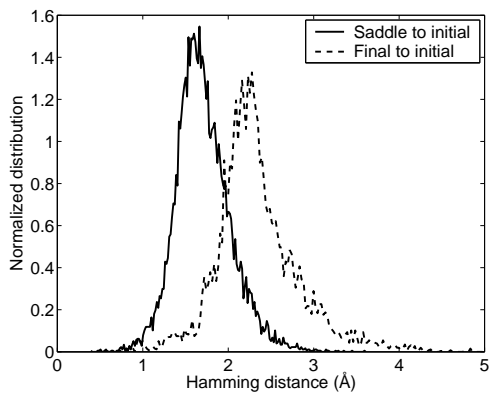}}

\caption{Distribution of the total displacement (Hamming distance) for
  the all events starting from Minimum 1 at the saddle point (solid
  line) and at the final minimum (dashed line). The distribution of
  Hamming displacements for unique events, as defined in the text, is
  essentially identical to this one.}

\label{fig:hamming}
\end{figure}

While the energy distribution inform us about the barriers heights,
the distribution of Hamming distances between configurations
(Fig.~\ref{fig:hamming}) provide some insight as to the rearrangements
taking place.  As was found previously in {\it a}-Si and {\it
  v}-SiO$_2$ most saddle points are somewhere at mid-way between the
initial and final stage.~\cite{mousseau00a,mousseau00b} This is
somewhat unexpected in the energy landscape picture: in a
high-dimensional space, any two randomly selected direction are
orthogonal. For truly high-dimensional events, therefore, one would
expect to find little correlation between the displacement at the
saddle point and that at the final minimum. This is not the case,
however, if events are taking place in a much restricted
sub-space. The distribution of displacements indicates therefore that
even though the simulation system is embedded in a 3000-dimensional
space, the effective sub-space in which each event takes place is much
smaller and local in nature; one must therefore be cautious when
interpreting the dynamics of a material based solely on the energy
landscape picture.

\begin{figure}
\centerline{\includegraphics[width=6.5cm]{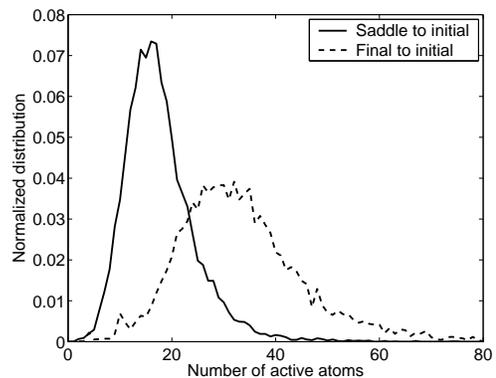}}

\caption{ Distribution of the number of activated atoms at the saddle
  point (solid line) and the final minimum (dashed line) for all
  events starting at minimum 1. The distribution is almost identical
  when only unique events, as defined in the text, are taken into
  account.}

\label{fig:number}
\end{figure}

\begin{figure}
\centerline{\includegraphics[width=6.5cm]{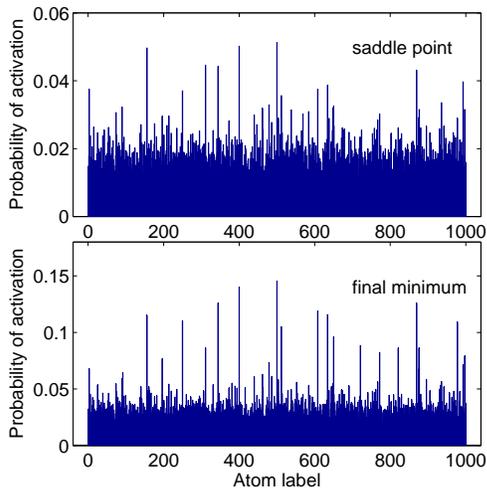}}
\caption{Probability of participating in a event for each atom of the
  model, computed using on all generated events. Top: probability
  of being an active atom at the saddle point; bottom: probability of
  being an active atom at the final minimum.} 
\label{fig:activity}
\end{figure}

The local nature of the dynamics is also found plotting the number of
active atoms participating in each event.  (Fig.~\ref{fig:number}). An
active atom is defined as one that has moved by more than 0.1 \AA\,
from its initial state.  As discussed in
Refs. \onlinecite{mousseau98,mousseau00a}, the value of the threshold
is chosen to be close to the typical atomic displacement due to
thermal vibration at room temperature.  We see that most events
involve a maximum of 25 atoms at the saddle point, and 40 at the final
minimum. If the threshold is increased to 0.4 or 0.5 \AA, the number
of active events drops typically to between 2 and 4 per event.  The
relatively narrow distribution of active atoms is not an artifact due
to some bias in the sampling of the cell.  Fig.~\ref{fig:activity}
shows the probability of each atom to participate into an event.  The
remarkably homogeneous figure confirms that the rearrangements can
take place anywhere in the network, with a participation probability
varying by at most a factor 3.

\subsection{Properties of the landscape}

\begin{figure}
\centerline{\includegraphics[width=6.5cm]{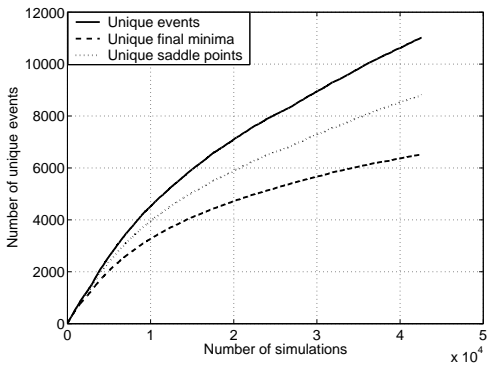}}
\caption{Solid line: Number of unique events as a function of the
  number of events sampled. The dashed and the dotted lines show the
  number of unique saddle points and final minima as a function of the
  number of events sampled. }
\label{fig:unique}
\end{figure}

As the sampling of events is random, the total data set contains
redundant events that might bias the analysis of the landscape. It is
therefore useful to look at the properties of ``unique'' events, i.e.,
to analyze the set of single copies of all events.  We identify each
event based on the identity of the participating atoms as well as on
the energy at the barrier and at the new minimum.  In order to
decrease the impact of the finite precision in the convergence at the
saddle and the minimum, we define the participating atoms as those
moving by at least $\delta r_{\mathrm{threshold}} = 0.4$ and 0.2 \AA,
respectively, at the saddle point and the final minimum. Two events
are considered identical if the active atoms are the same and the
energy barrier differs by less than 0.2 eV. We verified that the
precise values of the various thresholds do not affect the qualitative
results although the precise number of different events obviously
depends on the threshold. The number of unique events, as a function
of the number of events already sampled, is shown in
Fig. ~\ref{fig:unique}. We find 6519 different minima, 8799 different
saddles and 11014 different events generated within the 42 000+
sequence. On average, each event is therefore visited almost 4 times.

\begin{figure}
\centerline{\includegraphics[width=6.5cm]{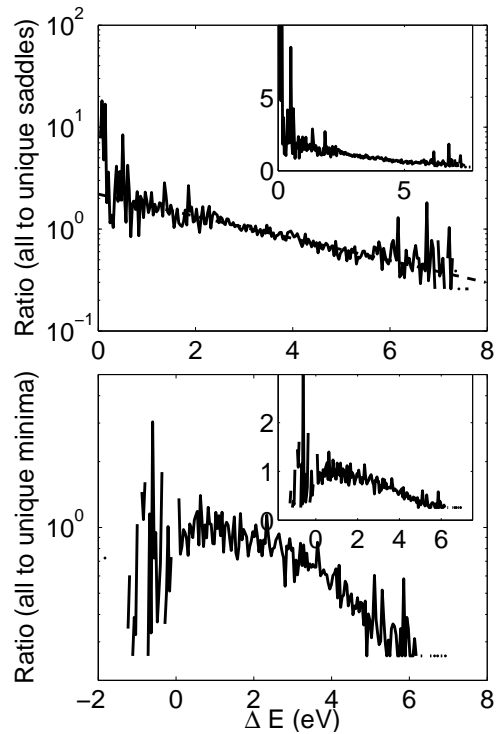}}
\caption{Top: Log-normal ratio of all saddles points generated from minimum 1
  over the unique ones only. A histogram for both distributions is
  first constructed, as a function of the energy, and the ratio is
  taken over this histogram. Bottom: same but the configurations at
  the final minimum.  The insets in both figures shows the same
  distributions plotted linearly.}
\label{fig:ratio}
\end{figure}

Using the distribution of unique events, it is possible to assess the
biases in the ART sampling. Fig.~\ref{fig:ratio} shows the ratio of
all saddle points or minima generated over the list of all unique
ones. As was the case for the Lennard-Jones clusters, a system with a
totally different energy landscape, ART sampling seems to select
events with an exponential bias, $\exp(- \Delta E/4)$.

These results are obtained using a subset of all events surrounding
Minimum 1. Even after 42000 events, the number of different events
around minimum is not yet converged. It is possible to obtain some
very crude estimate of this number. We first set the full distribution
of saddle points identical to that shown in
Fig. ~\ref{fig:unique}. Supposing a random selection of events, with
an exponential bias on the barrier height such as that of the previous
paragraph, taken from the distribution of unique events, we find that
growth of the total of unique events as a function of trial can be
reproduced with an exponential bias $\exp(-\Delta E/A)$ with $A$
between $0.40$ and $0.60$, and a total number of events somewhere
between 30,000 and 60,000 or 30 to 60 different saddle points per
atom.

\subsection{Entropy}

Having computed the distribution of activation energy barriers around
a minimum, it would be possible to consider associating a time scale
with the events. This can be done within the framework provided by
the transition state theory~\cite{glasstone41}. For activated
mechanisms, the diffusion rate is given by 
\begin{equation}
D = \frac{g}{2\alpha} \xi l^2 \nu_0 e^{-\Delta F /k_bT}
\end{equation}
where $g$ is the number of equivalent diffusion paths, $\alpha$
the spatial dimension, $l$ the length of the jump, $\nu_0$ a phonon
frequency, and $\Delta F = \Delta E - T\Delta S$, the variation in
the Helmholtz free energy between the transition state and the
initial state. 

In order to use the above equation, we need to measure the change in
entropy from the minimum to the transition state. We can do that using
the harmonic approximation:
\begin{equation}
\Delta S = k_b \ln \left[ \frac{ \displaystyle \prod_{i=1}^{3N}
    \nu_i^{(i)}} {\displaystyle \nu_0 \prod_{i=1}^{3N-1} \nu_i^{(s)}}\right]
\end{equation}
where $\nu_i^{(i)}$ and $\nu_i^{(s)}$ are the real phonon frequencies 
at the minimum and at the saddle point respectively. Since there are
$3N-1$ real frequencies at the transition state, we replace the
imaginary frequency by a typical phonon frequency, $\nu_0$. 

Assessing the time scale associated with leaving a given minimum
remains a very expensive task even within the harmonic approximation
as it would be necessary to compute the entropy associated with each
of the 30,000 to 60,000 saddle points.  A simpler approach poses that
the entropic prefactor is independent of the specific event.

\begin{figure}
\centerline{\includegraphics[width=6.5cm]{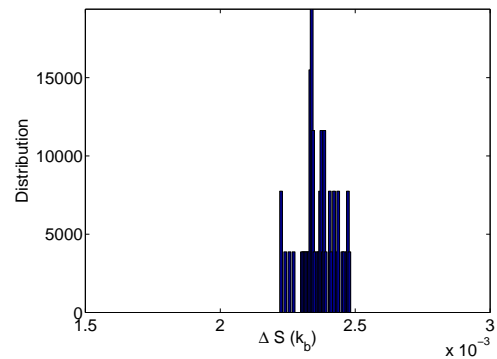}}

\caption{Distribution of the entropy barrier at the saddle point
  evaluated in the harmonic limit, for 50 different events selected at
random.}
\label{fig:entropy}
\end{figure}

We can ascertain the validity of such assumption by computing the
distribution of the entropic difference from the initial minimum to
the saddle point.  Due to the high cost of diagonalizing a $3000
\times 3000$ matrix, we have applied this procedure to 50
randomly-selected events.  As seen in Fig. ~\ref{fig:entropy}, the
contribution of the entropy to the activation is of the order of
0.0024 eV/K, with a variation of about $1.5\times 10^{-4}$ eV/K.
The fluctuations in the entropic
contributions to the barrier can therefore affect the attempt
frequency by about a factor 5. To a first
approximation, it is therefore reasonable to consider that the
entropy is only a multiplicative constant in the dynamics of the
system.

\subsection{Topological classification}

In the previous sub-sections, we have analyzed the properties of the
energy landscape itself. We now turn to the classification of the
events generated around a single minimum.  The topological
classification used here is described in detail in
Ref. ~\onlinecite{mousseau98}. Briefly, it only considers changes in
the coordination of the atoms, obtained through bond-breaking and
binding. Analyzing the more than 40 000 events generated here, we find
more than 2000 different types of topological events. Of those, we
discuss below only the events that occurred more than 1 \% of the time;
about 400 times in our sequence. Only 5 types of event satisfy this
criterion.

\begin{figure}
\vspace{5cm}

\centerline{\includegraphics[width=7.5cm]{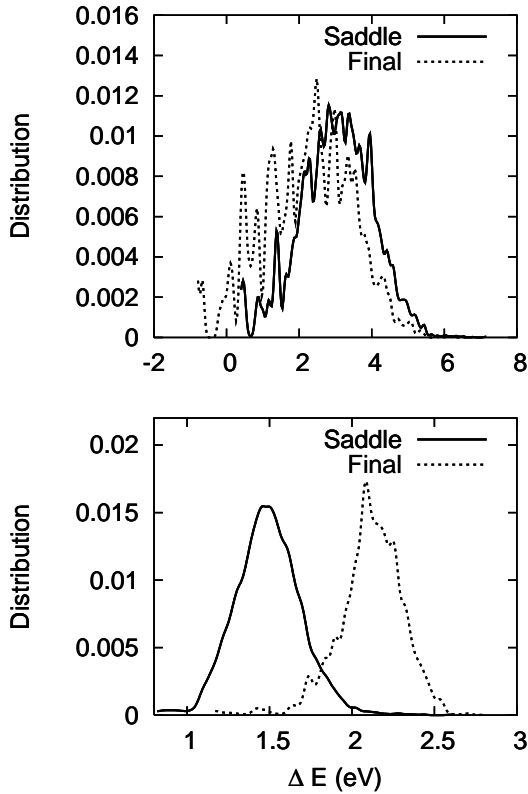}}

\caption{ 
Top: Energy distribution for the 8000 WWW-type events. Solid
  line: distribution at the saddle point; dotted line: distribution of
  the asymmetry energy.  Bottom: Distribution
  of the Hamming distance to the initial minimum at the saddle point
  (solid line) and the final minimum (dashed line).}
\label{fig:www}
\end{figure}

As in previous simulations, the most common event we find,
representing about 20 \% of all generated events, is the
Wooten-Winer-Weaire bond exchange mechanism (or abacbd, in our
notation), introduced almost 20 years ago in the sillium model. Since
then, this mechanism has been seen both in crystalline and amorphous
materials.
~\cite{wooten85,stillinger85,motooka96,tang97,bernstein00,goedecker02}
Figure~\ref{fig:www} shows the energy and Hamming distance
distribution for this mechanism.  Because of the frequency of this
mechanism, it might not be surprising to find that both the energy
and Hamming distance distributions follow closely those obtained for
the whole set of events. These distributions underline once more,
nevertheless, the r\^ole of strain in determining the activation
barrier for a given mechanism: the same topological jump can lead to
a new configuration with an energy varying by as much as 6 eV.

\begin{figure}
\centerline{\includegraphics[width=6.5cm]{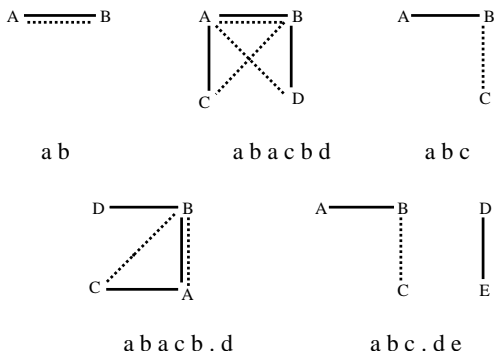}}

\caption{The 5 most common types of events, according to our
  topological classification. The full lines represent a bond present
  in the initial configuration and the dotted lines a bond present in
  the final configuration.}
\label{fig:representation}
\end{figure}

The topological nature of the other 4 dominant events is also shown
in Fig.~\ref{fig:representation}. The class $abc$ is associated with a
bond jumping from on pair of atoms to the next. $ab$ represents a bond
breaking at the saddle point and reforming. It does not involve any
topological change and is responsible for the low-energy peak in the
asymmetry energy distribution.  As such it is not very
interesting. Finally, the two other classes are modifications on the
three most common types.

\begin{table}
\begin{tabular}{cccc}
Class & Event Number & $\Delta E_{\mathrm{saddle}}$ & $\Delta
E_{\mathrm{final}}$ \\ \hline
abacbd &8478 & 3.04 & 2.26\\
abc &759 & 1.64 & 1.00\\
ab &735 & 3.04 & 0.21 \\
abacb.d &640 &2.47  &1.91 \\
abc.de &404 & 1.64 & 1.00\\
\end{tabular}
\end{table}

\section{Discussion and conclusions}
\label{sec:discussion}

The purpose of this study was to characterize in detail the energy
landscape around a single minimum in a well-relaxed model of {\it
  a}-Si.  Previous work has examined much shorter sequences of events
as the configuration relaxed and the atoms diffused.  Here, we
performed two extensive simulations, always starting from the same
two minima. Our results indicate that events are essentially local
with a barrier height limited by the cost of breaking one single
bond. We also found that there is very little variation in the
entropic barrier in spite of the wide spread in the energy part. This
is likely related to the fact that the overall shape of the energy
landscape is not finely dependent on the specifics of a
configuration. Finally, the bias found in ART nouveau indicates that
it might be possible to contemplate setting up a kinetic Monte Carlo
simulation on this system. 

It is interesting to note the strong dissimilarities between the
activation barrier distribution as computed from the initial and the
final minimum.  In particular, the barrier measured from the latter are
much lower that those measured from the former configuration. This
difference can be explained by the fact that the initial configuration
selected is very-well relaxed; most low energy barrier, associated
with a very unstable direction, would have been crossed during the
relaxation, in previous events. These states are likely not
contributing to diffusion or the relaxation of the network since they
would rapidly relax back into the initial minimum. The physically
relevant barrier distribution for defining the evolution of the
system is therefore that measured from the initial minimum. 

In spite of the 40,000+ events generated, the total number of events
around a minimum remains unknown. We can nevertheless estimate that
the number of different paths is somewhere between 30 and 60
paths/atom; as shown here, this number should be independent of the
size of the network as all events are local in nature.

Finally the stability of the distribution from one minimum to another,
as well as the narrow distribution of entropic barriers suggests that
it would be possible to develop an accelerated algorithm for {\it
  a}--Si, similar to that proposed by Henkelman and J\'onsson for the
diffusion of Cu on Cu~\cite{henkelman01}, and that of Hernandez-Rojas
and Wales for LJ glasses.~\cite{hernandez02} This is made even easier
by the exponential bias of ART, also seen in LJ clusters. The origin
of this bias is not understood but it simplifies significantly the
statistical analysis of an ART sampling.

At this point, however, the main limitation of this type of
simulation is the absence of detailed experimental results. 
Although the overall energy scale of the barriers is in agreement with
experimental numbers, it is difficult to assess whether or not the
theoretical description is correct: there is no information available
experimentally on the shape of the barrier distribution. It is
currently impossible to generate a better comparison with
experiment. Further experimental work is clearly needed 
to expand on the current available data and help ensure that the
advances in numerical results are on solid grounds.

\section{Acknowledgments}

We thank F. van Wijland for providing us with the solution do the
classical occupancy problem. We also thank D. Wales for useful
discussions and for providing their original data. FV is grateful to
the Natural Sciences and Engineering Research Council of Canada
(NSERC) for a summer scholarship. This work was partially supported by
NSERC and the Fonds Nature et Technologie du Qu\'ebec (NATEQ). NM is a
Cottrell Scholar of the Research Corporation.


\end{document}